\newcommand{\beq}{\begin{equation}}
\newcommand{\eeq}{\end{equation}}
\newcommand{\beeq}{\begin{eqnarray}}
\newcommand{\eeeq}{\end{eqnarray}}

\documentstyle[12pt,epsfig] {article}

\begin{document}

\begin{center}\bf
NEW COMPOSITE SUPERCONFORMAL STRING MODEL WITH TWO SCALES AND UNIFICATION OF FUNDAMENTAL INTERACTIONS  

\bigskip
V. A.  Kudryavtsev  \\  
    Petersburg  Nuclear Physics Institute\\
\bigskip
A b s t r a c t
\end{center}

 A  new approach to composite superconformal strings is considered .  This composite string model has two scales: first one  (~1 Gev) is for edging surfaces  and  second one (Planck scale)  is for ridge surfaces .  Nonlinear realization of two-dimensional superconformal  symmetry on edging  surfaces leads to superconvergence of one-loop planar diagrams. Some features of hadron amplitudes and  of hadron spectrum are discussed. Spectrum of closed strings  and a possibility of unified theory of fundamental interactions on the basis of this model are considered    
       
\newpage

\section {Introduction}

 In spite of significant success in treatment of  hadron interactions at high energies 
the quantum chromodynamics  (QCD) is  
unable up to now to give a consistent quantitative description of strong interactions at small and intermediate energies  (confinement problem).

        String models  (previously dual resonance models) \cite{1,2} had appeared as 
 a possible way  to describe soft strong  interactions.
Presently phenomelogical status of such approach seems to be even more impressive  than fifty years ago
since we have now stringlike spectrum of hadron states including not only leading Regge trajectories but  and
second and third daughter Regge trajectories for this spectrum \cite{3,4} up to spins equal to be
to 2 ,  3 and even up to 4 for mesons or  up to 11/2 for baryons .  However earlier  we  had not
consistent realistic string amplitudes for hadrons without negative norm states in physical spectrum and with
 intercept of leading meson ($\rho$) trajectory to be equal to one half \cite {5}. For all previous classical string models
 consistent amplitudes had required the  intercept of leading meson trajectory to be equal to one.   It turns out to be possible 
to build  a composite string model to be compatible with these requirements \cite{6} with
 intercept of leading meson ($\rho$) trajectory to be equal to one half. This model  gives 
 realistic description of the hadron spectrum and brings to  correct interaction of arbitrary even number of $\pi$ -mesons  which satisfies
the Adler-Weinberg condition for soft $\pi$ -mesons \cite{7}.  Composite superconformal  string model  provides  a generalization of well known four-point  Lovelace-Shapiro amplitude to  arbitrary N-point pion amplitudes.

           However this model \cite {6} faces problems when treating nucleons and a possible massless tensor state ( strong graviton) in
nonplanar one-loop  diagram. Namely  in classical string approach we have the strict connection between  slopes of Regge trajectories for open string states and for closed string states. This constraint  would be forced give up classical string description  for hadrons. It is proved to be possible to overcome  these  problems in a new  reformulated composite superconformal  string  model  which we shall consider  below.

            In this new composite string model we have as for  the previous composite string model  one basic  two-dimensional surface
and  two  additional edging  surfaces for mesons $(Q \overline{Q})$ and three additional  surfaces (two edging and one ridge surfaces) for baryons (QQQ)
  in accordance with dual quark diagrams (Harari -Rosner diagrams) \cite{8} .
                        In so doing quark lines  were replaced by above- mentioned  additional  two-dimensional   surfaces. Namely this topology brings to 
 composite string models \cite {6}.  
  For the  composite string to be considered we take  different scales for edging surfaces $(\sqrt{\alpha'_{H}})^{-1}$ (~1 Gev) and for ridge surfaces  $(\sqrt{\alpha'_{Pl}})^{-1}$ ($~  10^{19}$ Gev). 
As we shall see it allows  us to separate hadron states with usual hadron scale (1 Gev) in the open string sector from massless graviton state with Planck 
scale for Regge trajectories in the closed string sector.

           Second distinctive  feature of this model  in relation to the previous one \cite{6} is  a nonlinear realization of two-dimensional superconformal 
symmetry for  $ \Phi $ -fields  on edging surfaces  (see Antoniadis, Narain and others).  Namely this way leads us to superconverging one-loop planar
string  diagrams due to excess of number of fermion two-dimensional fields over number of boson ones. We  remind  that we have  not 
 supersymmetry in target space  and  hence this property of this  convergence  is very important for consistency of our approach.

 \section {Formulation of the model. Two-dimensional fields}

            For composite string as it is considered above  we use   the conventional basic two-dimensional surface  and additional two-dimensional surfaces which are reproducing  the  picture  of quark lines in dual quark  diagrams   by Harari and Rosner \cite{8}.

          So we have two edging surfaces in addition to the basic one for description of Q anti Q -mesons and  three additional surfaces (two edging and one
ridge surface )  for description of  QQQ baryons.
     The description of  composite string amplitudes in the framework of functional integration   proposes  free Polyakov actions 
for each two-dimensional   surface and   free two-dimensional  fields on it.  In so doing  for $\mu =0,1,2,3$ (Lorentz index) we have on  the basic  two-dimensional surface two- dimensional   fields $\partial {X^{\mu}}$   wiith superpartner fields $H^{\mu}$.
   In operator presentation we have  well known expressions for them:

\begin{eqnarray}
\partial X_\mu (z)&=&P_\mu +\sum_{n}a_{n\mu}z^{n};\;\;
 \\
\left[ a_{n\mu},a_{m\nu}\right]&=&-ng_{\mu\nu}
\delta_{n,-m}\\
H_\mu (z)&=&\sum_{r} b_{r\mu} z^{r} \\
\left\{ b_{r\mu}, b_{s\nu}\right\} &=&
-g_{\mu\nu}\delta_{r,-s}.
\end{eqnarray}
  Here  n,m are integer numbers, r,s are half-integer numbers. 

        In addition to these  fields $\partial {X^{\mu}}$ and superpartner fields $H^{\mu}$  for $\mu =0,1,2,3$  we have six  two- dimensional  fields 
 $I^{(a)}$( $a$=1,2,3,4,5,6 ) ( which have the conformal spin to be equal to one as for $\partial {X^{\mu}}$ ) \\
  and six  superpartner fields   $\Theta^{(a)}$  ( $a$=1,2,3,4,5,6 )
                                   with corresponding equations  for them
\begin{eqnarray}
I^{(a)} (z)&=&I^{(a)}_0 +\sum_{n}I^{(a)}_{n}z^{n};\;\;
 \\
\left[I^{(a)}_{n},I^{(b)}_{m}\right]&=&n\delta_{n,-m}\delta_{a,b}           \\
\Theta^{(a)} (z)&=&\sum_{r}\Theta^{(a)}_{r} z^{r} \\
\left\{\Theta^{(a)}_{r},\Theta^{(b)}_{s}\right\} &=&\delta_{r,-s}\delta_{a,b}
\end{eqnarray}
New   fields  give us  necessary  dependence on quantum numbers (isospin, flavor and so on).

    Namely these ten fields ( $\partial {X^{\mu}}$ with superpartner field $H^{\mu}$  for $\mu =0,1,2,3$ and  $I^{(a)}$( $a$=1,2,3,4,5,6 ) with   superpartner fields   $\Theta^{(a)}$     instead of usual $\partial {X^{\mu}}$ with superpartner field $H^{\mu}$  for $\mu =0,1,...9$  provide necessary critical number of two- dimensional  fields for appearence of closed string states from nonplanar one-loop diagrams.

            In correspondence with quark dual diagrams  we have introduced $\lambda_{\alpha}$ operators to carry quark flavors and quark spin degrees of freedom. As we shall see that products of operators  $\lambda $  (as $\overline{\lambda}_i  \tau^{(b)}\gamma_{5}\lambda_{i+1})$ for example   will be  eigenvectors of the operators $I_0^{(a)}$ and therefore these products  will be  an analog of $\exp(ik_iX_0)$ for field $\partial  X$ and $\exp(ik_i\tilde Y_0^{(i)})$ for field $Y^{(i)}$,  as $\hat p_i(\exp(ik_iX_0))=k_i(\exp(ik_iX_0))$ and $Y_0^{(i)}(\exp(ik_i\tilde Y_0^{(i)}))=k_i(\exp(ik_i\tilde Y_0^{(i)}))$ .( Here  $\tau^{(b)}$ are usual Pauli matrices for isospin).     This approach replaces usual transition to extra dimensions and allows introduce the quark quantum numbers in natural way.
 In addition,  we obtain an attractive interpretation of the Chan-Paton factor .
\begin{eqnarray}
\langle0|\lambda^{+}=0,\\ \lambda |0\rangle=0.
\end{eqnarray}
\begin{eqnarray}
\left\{\bar\lambda_{\alpha},\lambda_{\beta} \right\}=\delta_{\alpha,\beta},\\\bar\lambda=\lambda^{+} T_0,\\T_0=\gamma_0\otimes\tau_2;\\
\tilde\lambda=\lambda T_0
\end{eqnarray}

  Further we define charges $I^{(a)}_0$ for these fields $I^{(a)}$( $a$=1,2,3,4,5,6 ) with the exception of   $I^{(2)}$  as  the sums of products of isotopic generators  $T^{(l)},l=1,2,3$  (compare with  the  Heisenberg  hamiltonian for spin chains  $\sum_i \sum_{l=1,2,3} S^{(l)}_i S^{(l)}_{ i+1}$,
 where i  is a number in the chain ) .

 The explicit expression for  $ I^{(2)}_{0}$  will be the linear sum of isotopic generators.

      These charges shoulld be  Lorentz invariant  and  be commutating each one with all others. This choice is a very limited one. It leads to inclusion of all three
generations of quarks in order to obtain   the critical value to be equal to 10  for the number of currents including  $\partial {X^{\mu}}$   for $\mu =0,1,2,3 $ and  $I^{(a)}$, a=1,2,3,4,5,6.

Let us to write the explicit expression for  $ I^{(1)}_{0} $  :
\begin{eqnarray}
 I^{(1)}_0=  g_{(1)} \sum_{l=1,2,3} \sum_{\mu=0,1,2,3}((\sum_{i,j}  (T^{(l)}_{j \mu} T^{(l) \mu}_{i }) + g_{ed}(\sum_{i_1,i_2}
 (T^{(l) \mu}_{i_1}T^{(l )}_{i_2 \mu}) )
\end{eqnarray}

\begin{eqnarray}
T^{(l)}_{\mu j}=  [\frac 1{2}\bar\lambda^{(+)}_{j}  \tau^{(l)} \gamma _{\mu}\lambda_{j}]  ;      
T^{(l) \mu}_{i}= [\frac 1{2}\bar\lambda^{(+)}_{i}  \tau^{(l)} \gamma ^{\mu}\lambda_{i}] ; 
\end{eqnarray}

This operator $ I^{(1)}_0$ is an isotopic invariant one. It is very  important  for  real  isotopic spins $ \sum_{l=1,2,3}T^{(l)}T^{(l)}$ dependence of hadron masses.

Here $i$ are numbers of edging surfaces  and $j$ are numbers of ridge  surfaces.

Let us to write the explicit expression for  $ I^{(2)}_{0} $ .  Only   for this  case
   we use  the  linear   sum  :

 \begin{eqnarray}
 I^{(2)}_0=g_{(2)}( \sum_{i}\tilde Q^{(edge)}_{i}+   \sum _{j}\tilde Q^{(ridge)}_{j})
\end{eqnarray}

As above  $i$ are numbers of edging surfaces  and $j$ are numbers of ridge  surfaces.
Here
\begin{eqnarray}
\tilde  Q^{(edge)}_{i}= [\frac 1{2}\bar\lambda^{(+)}_{i}\tau^{(3)}\lambda_{i}] ;       \cr
\tilde Q^{(ridge)}_{j}= [\frac 1{2}\bar\lambda^{(+)}_{j}(1+ \tau^{(3)})\lambda_{j}] ; 
\end{eqnarray}

This charge  $ I^{(2)}_0$  will be of importance in definition of a  string interpretation of electromagnetic  interaction due to corresponding 
closed string states to be appeared  in nonplanar one-loop string diagrams for this theory.

and   for $ I^{(3)}_0$:

\begin{eqnarray}
 I^{(3)}_0=g_{(3)} \sum_{\mu=0,1,2,3}( \sum_{j_1,j_2}\tilde Q_{\mu j_1}\tilde Q^{\mu}_{j_2})
\end{eqnarray}

As above  $j$ are numbers of ridge  surfaces.
Here
\begin{eqnarray}
 \tilde Q_{\mu j}= [\frac 1{2}\bar\lambda^{(+)}_{j}\tau^{(3)}\gamma_{\mu}\lambda_{j}] ;      \tilde Q^{\mu}_{j}= [\frac 1{2}\bar\lambda^{(+)}_{j}\tau^{(3)} \gamma^{\mu}\lambda_{j}] ; 
\end{eqnarray}

 And we have similar forms of charges   $ I^{(a)}_0$  for  a=4,5,6 ; $ a-3=a_f= 1,2,3 $  is a flavor index for three  generations of quarks:

\begin{eqnarray}
 I^{(a)}_0=T^{(+)}_{\mu(a_f)(c_f)j_1} T^{\mu(-)}_{(c_f)(a_f)j_2} 
\end{eqnarray}  
Here 
\begin{eqnarray}
T^{(+)}_{\mu(a_f) (c_f)j}= [\frac 1{2}\bar\lambda^{(+)}_{ (a_f)j}(1-\gamma_5)\gamma_{\mu} M_{(a_f)( c_f)}  \tau^{(+)}\lambda_{ (c_f)j}];\\
T^{\mu (-)}_{ (c_f)(a_f)j}= [\frac 1{2}\bar\lambda^{(+)}_{ (c_f)j}(1-\gamma_5)\gamma^{\mu}  M_{(c_f)(a_f)}^*   \tau^{(-)}\lambda_{(a_f)j}];\\
\end{eqnarray}

$M_{(a_f)( c_f)};M_{(c_f)(a_f)}^*$   are  elements of the CKM matrix M for $( c_f)$-th row  and $ (a_f)$-th column correspondingly; $(a_f),(c_f)$ are flavor indices.

             For  edging and ridge surfaces we have  fields  which carry momentum  to be corresponding to  these  additional   surfaces . 
 So for $\mu =0,1,2,3$ (Lorentz index)  we  have $ Y ^{\mu}_{(i)}$, $ f  ^{\mu}_{(i)}$(superpartner) and $ Y ^{\mu}_{(j)}$, $ f  ^{\mu}_{(j)}$(superpartner)    fields for i-th edging or for j-th ridge  surfaces   instead of $\partial {X^{\mu}}$  with superpartner field $H^{\mu}$  for the base  surface .
  Let us remind that we have two scales for edging and ridge surfaces here.

\begin{eqnarray}
 Y^{(i)}_{\mu} (z)&=&\sqrt{\alpha'_{H}}Y^{(i)}_{0\mu} +\sum_{n\neq 0}Y^{(i)}_{n\mu}z^{n};\;\; \\
 Y^{(j)}_{\mu} (z)&=&\sqrt{\alpha'_{Pl}}Y^{(j)}_{0\mu} +\sum_{n\neq 0}Y^{(j)}_{n\mu}z^{n};\;\;\\
P_{\mu} &=&\sum_i\sqrt{\alpha'_{H}}Y^{(i)}_{0\mu}+\sum_j \sqrt{\alpha'_{Pl}}Y^{(j)}_{0\mu}
\end{eqnarray}
Here i   is a number of an edging surface   and    j is a number of a ridge surface; 
Eigenvalues of $P_\mu = \sum_i\sqrt{\alpha'_{H}}k^{(i)}_{\mu}+\sum_j \sqrt{\alpha'_{Pl}}k^{(j)}_{\mu}$.    These definitions are   in correspondence 
with previous considerations.

             For simple cases   the sums are including  two edging surfaces ( for Q anti Q   channel )  and three surfaces
 (two edging and one ridge surface )  for QQQ channel.   
            
          As distiguished from the  previous  model   we take 
some triplets of anticommutating  fields  of conformal spin to be equal to one half $\Phi_A^{(i)}$, $\Phi_B^{(i)}$, $\Phi_C^{(i)}$    for each i-th edging surface
 instead  of a single field $ J ^{(i)}$(conformal spin to be equal to 1)  with superpartner $\Phi ^{(i)}$(conformal spin to be equal to one half) to be used  in previous version .
  Thus  we can  introduce nonlinear realization of superconformal symmetry 
on these surfaces.

   We have
\begin{eqnarray}
\Phi_A  ^{(i)} (z)&=&\sum_{r}\Phi^{(i)}_{(A) r} z^{r} \\
\left\{\Phi^{(i)}_{(A) r} ,\Phi^{(i)}_{(A)s} \right\} &=&\delta_{r,-s}
\end{eqnarray}
and the same equations for $\Phi^{(i)} _{(B)}, \Phi^{(i)}_{(C)}$ .

     In operator formalism we  can present  Virasoro superalgebra generators as follows 

\begin{eqnarray}
G_r=G_r^{Lor}+ G_r^{Int},
\end{eqnarray}

\begin{eqnarray}
G_r^{Lor}= \frac{1}{2\pi}  \int_0^{2\pi} d\tau  [\sum_{\mu}[\partial X_{\mu}H^{\mu}+\sum_{i,\mu} Y_{\mu}^{(i)}f^{(i)\mu}]e^{-ir\tau},\\
G_r^{Int}= \frac{1}{2\pi}  \int_0^{2\pi} d\tau  [\sum_{\nu}(I^{\nu}\Theta^{\nu}+\sum_{i} \Phi^{(i)}_A  \Phi^{(i)}_B \Phi^{(i)}_C)] e^{-ir\tau}
\end{eqnarray}

\section { Nucleon vertices}

      As before in previous approach we shall formulate  basic  vertices for ground states . Namely these basic states together with  two-dimensional fields for accepted topology  define all physical amplitudes of our model. In Koba-Nielsen representation  N-string Born amplitudes are given as integrals of
vacuum expectation of  N basic vertices product over z-circle (or z-axis) for open strings.

\begin{eqnarray}
A_N = \int\prod dz_i\langle0|\hat{V}_1(z_1)\hat{V}_2(z_2)
\hat{V}_3(z_3)...\hat{V}_{N-1}(z_{N-1})\hat{V}_N(z_N)|0\rangle
\end{eqnarray}
\begin{eqnarray}
\hat{V}_i(z_i)\ =\ z^{-L_0}_i \hat{V}_i(1) z^{L_0}_i.
\end{eqnarray}

             These vertices $\hat{V}_i$ have the well-known expressions for the  classical  Neveu--Schwarz model:
\begin{eqnarray}
\hat{V}_i(z_i)\ =\ z_i^{-L_0}\left[G_r,:\exp{ip_i X(1)}:\right]z_i^{L_0}, \cr
 :\exp{(i p_{i}X(1))} :\ =\ \exp {(i p_i X^{(+)}(1))}\ \exp{(ip_i X_0 )} \exp{(ip_i X^{(-)}(1))},
\end{eqnarray}

\begin{eqnarray}
G_r^{NS} = \frac1{2\pi}\int\limits_0^{2\pi}d\tau\left(H^{\mu}\frac d{d\tau} X_\mu+ \hat{P}_\nu H^\nu\right)e^{-ir\tau}
\end{eqnarray}

\begin{eqnarray}
\hat{V}_i(1)= (p_i H(1)):\exp{(ip_i X(1))}\ :\equiv\cr \equiv(\sum_r p_ib_r)\exp{(ip_i X^{(+)}(1))}\ \exp{(ip_i X_0)}\ \exp{(ip_i X^{(-)}(1))}\ \equiv \cr
\equiv (\sum_r p_ib_r)\exp{(-ip_i\sum_n \frac {a_{-n}}{n})}\ \exp{(ip_i
X_0 )}\  \exp{(ip_i \sum_n \frac {a_{n}}{n})}.
\end{eqnarray}

            If we go to the new composite strings operator vertices   we have to take into account new  two-dimensional fields. 
 Then the vertex operator $\hat V_{i,i+1}$ for $\pi$-meson emission has assumed  in  the following form :
\begin{eqnarray}
\hat V_{i,i+1}(z_i)=z_i^{-L_0} \left[ G_r,\hat W_{i,i+1} \right] z_i^{L_0}, \\
\hat W_{i,i+1} = \hat R_i^{out}\hat R_{NS}\hat R_{i+1}^{in}.
\end{eqnarray}
The operators $\hat R_i^{out}$ and $R_{i+1}^{in}$ are defined by fields on $i$-th and $(i\!+\!1)$-th edging surfaces. The operator $\hat R_{NS}$ is defined by fields on the basic surface. They have the  same structure as the operator vertices of old classical Neveu--Schwarz string model $\hat{V}_i(z_i)$ :

      Namely  we  have
\begin{eqnarray}
\hat R_i^{out}=\sum_{E=A,B,C}\exp(\xi_i\sum_n\frac{J^{(i)}_{(E)-n}}{n})\exp(\sqrt{\alpha'_{H}}k_i\sum_n\frac{Y^{(i)}_{-n}}{n})\cr 
\exp(ik_i\tilde Y_0^{(i)})
 \exp(-\sqrt{\alpha'_{H}}k_i\sum_n\frac{Y^{(i)}_{n}}{n})\sum_{E=A,B,C}\exp(-\xi_i\sum_n\frac{J^{(i)}_{(E)n}}{n}),
\end{eqnarray}
Here $J^{(i)} _{(A)}=\Phi^{(i)} _{(B)} \Phi^{(i)}_{(C)};J^{(i)} _{(B)}=\Phi^{(i)} _{(C)} \Phi^{(i)}_{(A)};\\
J^{(i)} _{(C)}=\Phi^{(i)} _{(A)} \Phi^{(i)}_{(B)}$
\begin{eqnarray}
R_{i+1}^{in}=\sum_{E=A,B,C}\exp(-\xi_{i+1}\sum_n\frac{J^{(i+1)}_{(E)-n}}{n})\exp(-\sqrt{\alpha'_{H}}k_{i+1}\sum_n\frac{Y^{(i+1)}_{-n}}{n}) \cr \exp(-i\sqrt{\alpha'_{H}}k_{i+1}
\tilde Y_0^{(i+1)})  \exp(\sqrt{\alpha'_{H}}k_{i+1}\sum_n\frac{Y^{(i+1)}_{n}}{n})\sum_{E=A,B,C}\exp(\xi_{i+1}\sum_n\frac{J^{(i+1)}_{(E)n}}{n})
\end{eqnarray}
Here $J^{(i+1)} _{(A)}=\Phi^{(i+1)} _{(B)} \Phi^{(i+1)}_{(C)};J^{(i+1)} _{(B)}=\Phi^{(i+1)} _{(C)} \Phi^{(i+1)}_{(A)};\\
J^{(i+1)} _{(C)}=\Phi^{(i+1)} _{(A)} \Phi^{(i+1)}_{(B)}$
\begin{eqnarray}
\hat R_{NS}=\exp(-\sum_a \zeta^{(a)}\sum_n\frac{I^{(a)}_{-n}}{n})\exp(-p_{i,i+1}\sum_n\frac{a_{-n}}{n})
  \cr  [\bar\lambda^{(+)}_i  \tau^{(b)}\gamma_{5}\lambda^{(-)}_{i+1}] 
\exp(p_{i,i+1}\sum_n\frac{a_{n}}{n})\exp(\sum_a\zeta^{(a)}\sum_n\frac{I^{(a)}_{n}}{n})
\end{eqnarray}
 Here  (a =1,2,3,4,5,6 ), (b=1,2,3 is the isotopic index),\\

$ \zeta^{(a)} $ is an eigenvalue of  $ I_0 ^{(a)}$. \\

         For the pion wave function   we have
\begin{eqnarray}
\Psi_\pi= [\bar\lambda^{(+)}_i  \tau^{(b)}\gamma_{5}\lambda^{(-)}_{i+1}]
\end{eqnarray}
Also we  have used values:
\begin{eqnarray}
p^\mu_{i,i+1} = \sqrt{\alpha'_{H}}k_{(i)}^{\mu} - \sqrt{\alpha'_{H}}k_{(i+1)}^{\mu}  
\end{eqnarray}
It is in correspondence with  (27).

 So we have some relation between momenta (charges) flowing on the basic surface and on edging surfaces.
In this case  (for pions) ridge surfaces are absent.

We have to fulfill conditions for momenta: ${k_i^2=k_{i+1}^2=0}$\\
  to ensure $k_i^{\mu}  Y^{(i)}_{\mu} (z) ;k_j^{\mu}  Y^{(j)}_{\mu} (z)$ to be supercurrent conditions  in our model\\  (similarly  as for previous version    \cite {6}).

      In similar way  we can  build  the vertices  for emission of nucleons in this approach.\\
Since we have transitions $N\bar N -> \pi$ and the vertex $ V_{(\pi)}$  consists odd number of  anticommuting components of two-dimensional fields
(i.e.$ V_{(\pi)}$  is a vertex of negative G-parity)  nucleon vertices $V_{(N)}$  should be
   the sums of two components   of different values   of  G-parity in relation to odd or even number of anticommuting components of two-dimensional fields :

  \begin{eqnarray}
V_{(N)}=V_{(N)}^{(+)} + V_{(N)}^{(-)}.
\end{eqnarray}

 At first  we  define corresponding  wave functions of nucleons for $V_{(N)}^{(+)}$  and $ V_{(N)}^{(-)}$ : $\Psi_{(N)}^{(+)}$ and $ \Psi_{(N)}^{(-)}$:
      
 \begin{eqnarray}
 \Psi_{(N)}^{(+)}=B_{(+)}\lambda^{(+)}_{j}  (\tilde\lambda^{(+)}_i \lambda^{(+)}_{i+1}) 
\end{eqnarray}

\begin{eqnarray}
\Psi_{(N)}^{(-)}=  A_{(-)} \sum_{b=1,2,3}  \tau^{(b)}\gamma_{5}\lambda^{(+)}_{j}  (\tilde\lambda^{(+)}_i  \tau^{(b)}\gamma_{5}\lambda^{(+)}_{i+1})
\end{eqnarray}

      We  suppose   the positive parity for nucleons
\begin{eqnarray}
    \frac { \hat P_{(N)}} {m_N} \lambda^{(+)}_{j}= \lambda^{(+)}_{j};  P_{(N)}= p_{i,j,i+1}
\end{eqnarray}
 or equivalently
\begin{eqnarray}
 \gamma_0  \lambda^{(+)}_{j} =   \lambda^{(+)}_{j}
\end{eqnarray}
      Here i,i+1 are numbers of  edging surfaces , j is the number of the ridge surface.

      Then the vertex operator $\hat V_{i,j,i+1}^{(-)}$ of the negative G-parity for nucleon emission has assumed  the following form :
\begin{eqnarray}
\hat V_{i,j,i+1}^{(-)}(z_i)=z_i^{-L_0} \left[ G_r,\hat W_{i,j,i+1} \right] z_i^{L_0}, \\
\hat W_{i,j,i+1} = \hat W_{i,j}^{out}\hat W_{NS}\hat W_{i+1}^{in}.
\end{eqnarray}
The operators $\hat W_i^{out}$ and $W_{j,i+1}^{in}$ are defined by fields on $i$-th,$(i+1)$-th  edging surfaces and $(j)$-th ridge surface. The operator $\hat W_{NS}$ is defined by fields on the basic surface. 

      Namely  we  have
\begin{eqnarray}
\hat W_{i,j}^{out}=\sum_{E=A,B,C}\exp(\xi_i\sum_n\frac{J^{(i)}_{(E)-n}}{n}) \exp(\sqrt{\alpha'_{H}} k_i\sum_n\frac{Y^{(i)}_{-n}}{n})
\exp(\sqrt{\alpha'_{Pl}} k_j\sum_n\frac{Y^{(j)}_{-n}}{n})\cr 
\exp(i\sqrt{\alpha'_{H}}k_i\bar Y_0^{(i)})\exp(i\sqrt{\alpha'_{Pl}}k_j\bar Y_0^{(j)}) \cr
 \exp(-\sqrt{\alpha'_{H}}k_i\sum_n\frac{Y^{(i)}_{n}}{n})\exp(-\sqrt{\alpha'_{Pl}}k_i\sum_n\frac{Y^{(j)}_{n}}{n})
\sum_{E=A,B,C}\exp(-\xi_i\sum_n\frac{J^{(i)}_{(E)n}}{n}),
\end{eqnarray}
Here $J^{(i)} _{(A)}=\Phi^{(i)} _{(B)} \Phi^{(i)}_{(C)};J^{(i)} _{(B)}=\Phi^{(i)} _{(C)} \Phi^{(i)}_{(A)};\\
J^{(i)} _{(C)}=\Phi^{(i)} _{(A)} \Phi^{(i)}_{(B)}$
\begin{eqnarray}
W_{i+1}^{in}=\sum_{E=A,B,C}\exp(-\xi_{i+1}\sum_n\frac{J^{(i+1)}_{(E)-n}}{n})\exp(-\sqrt{\alpha'_{H}} k_{i+1}\sum_n\frac{Y^{(i+1)}_{-n}}{n}) \cr \exp(-i\sqrt{\alpha'_{H}}k_{i+1}
\bar Y_0^{(i+1)})  \exp(\sqrt{\alpha'_{H}}k_{i+1}\sum_n\frac{Y^{(i+1)}_{n}}{n})\sum_{E=A,B,C}\exp(\xi_{i+1}\sum_n\frac{J^{(i+1)}_{(E)n}}{n})
\end{eqnarray}

Here $J^{(i+1)} _{(A)}=\Phi^{(i+1)} _{(B)} \Phi^{(i+1)}_{(C)};J^{(i+1)} _{(B)}=\Phi^{(i+1)} _{(C)} \Phi^{(i+1)}_{(A)};\\
J^{(i+1)} _{(C)}=\Phi^{(i+1)} _{(A)} \Phi^{(i+1)}_{(B)}$
\begin{eqnarray}
\hat W_{NS}=\exp(-\sum_a \zeta^{(a)(-)}_{i,j,i+1}\sum_n\frac{I^{(a)}_{-n}}{n})\exp(-p_{i,j,i+1}\sum_n\frac{a_{-n}}{n})
   \cr  \Psi_{(N)}^{(-)} 
\exp(p_{i,j,i+1}\sum_n\frac{a_{n}}{n})\exp(\sum_a\zeta^{(a)(-)}_{i,j,i+1}\sum_n\frac{I^{(a)}_{n}}{n})
\end{eqnarray}

 Here  (a =1,2,3,4,5,6 ), (b=1,2,3 is an isotopic index),    
$ \zeta^{(a)(-)}_{i,j,i+1}$ is an eigenvalue of  $ I_0 ^{(a)}$ 
 for the component of nucleon wave function  $ \Psi_{(N)}^{(-)}$.
 \begin{eqnarray}
\Psi_{(N)}^{(-)}=  A_{(-)}(\frac1{2} (1- \frac { \hat P_{(N)}} {m_N}) \sum_{b=1,2,3}  \tau^{(b)}\gamma_{5}
 \lambda^{(+)}_{j})  (\tilde\lambda^{(+)}_i  \tau^{(b)}\gamma_{5}\lambda^{(+)}_{i+1}
\end{eqnarray}

      The   vertex operator $\hat V_{i,j,i+1}^{(+)}$ of the positive G-parity for nucleon emission
 has  the structure of the Bardarkci-Halpern vertex operator \cite{9}
with an additive factor $ \hat F$ which brings this expression to even number of  anticommuting components of two-dimensional fields:

\begin{eqnarray}
\hat V_{i,j,i+1}^{(+)}(z_i)=z_i^{-L_0} \left[ G_r,\hat W_{i,j,i+1}^{(+)} \right] z_i^{L_0}, \\
\hat W_{i,j,i+1}^{(+)} = \hat W_{i,j}^{out}\hat F(z=1)\hat W_{NS}^{(+)}\hat W_{i+1}^{in}.
\end{eqnarray}

\begin{eqnarray}
\hat W_{NS}^{(+)}=\exp(-\sum_a \zeta^{(a)(+)}_{i,j,i+1}\sum_n\frac{I^{(a)}_{-n}}{n})\exp(-p_{i,j,i+1}\sum_n\frac{a_{-n}}{n})
\exp(-ip_{i,j,i+1}X_0)   \cr  \Psi_{(N)}^{(+)} 
\exp(p_{i,j,i+1}\sum_n\frac{a_{n}}{n})\exp(\sum_a\zeta^{(a)(+)}_{i,j,i+1}\sum_n\frac{I^{(a)}_{n}}{n})
\end{eqnarray}
$ \zeta^{(a)(+)}_{i,j,i+1}$ is an eigenvalue of  $ I_0 ^{(a)}$ 
 for the component $ \Psi_{(N)}^{(+)}$ of nucleon wave function.
\begin{eqnarray}
 \Psi_{(N)}^{(+)}=B_{+)}(\frac1{2}( \frac { \hat P_{(N)}} {m_N} +1) \lambda^{(+)}_{j})  (\tilde\lambda^{(+)}_i \lambda^{(+)}_{i+1}) 
\end{eqnarray} 

\begin{eqnarray}
\hat F=f_1(p_{i,j,i+1} H )+\sum_a f^{(a)}_2\Theta^{(a)}
\end{eqnarray}

\section {Closed composite string states}

                   As  it was considered  above   planar one-loop diagrams  are superconvergent ones  here  due to excess of number of  anticommutating modes over number of commutating ones from nonlinear realization of conformal supersymmetry on edging surfaces (triplets of anticommutating  $ \Phi_A,\Phi_B,\Phi_C $ fields  instead  of the  pairs of  the commutating J  and the anticommutating  $ \Phi$ -fields  in usual linear realization of conformal supersymmetry
 as in previous version (6) ). 

   It does not work  for nonplanar  one-loop diagrams where there is  not   summing  over edging  surface fields . And these  nonplanar  
one-loop diagrams bring to appearance of  closed string states in the critical case (10 fields of conformal spin equal to one) as before in the  Neveu-Schwarz 
theory. 
    It is worth to  be noted that  it  is required two (or more than two) ridge surfaces on the tube under consideration   for  appearance of  these closed string states
 in the critical case  in order to have a  nonvanished momentum  for this  tube (see (27) .  It is  evidently there are no edging surfaces for closed strings.
 So  we have  only  the Planckean scale  $(\sqrt{\alpha'_{Pl}})^{-1}$ ($ \approx  10^{19}$ Gev) for all masses of closed string states  except ground states.
Masses of closed ground states can be determined  by $g_{(2)},  g_{(3)}$  and   $g_{(W)}$ constants  in our currents   (see (16),(18) and (20)) which  determine
 a string interpretation of electroweak interactions in our model.
 Since we have $ m_{W} \approx 80$  Gev  and   $ m_{Z} \approx 91$  Gev  and Planck scale $ \approx  10^{19}$ Gev our constants  $g_{(2)},  g_{(3)}$ would  be to $\approx   10^{-17}$. As it was mentioned above  the isotopic invariant  current $ I^{(1)}$  plays  an  important role for hadron spectrum and $ I^{(1)}_0=        g_{(1)}; 1> g_{(1)} > \frac 1{2} $.
      It is  noteworthy that it  is possible here to obtain a string interpretation for leptons as closed string states  with only one ridge surface on the tube.
More detailed analysis  would be performed  in  further  publications.

      I would like to thank  participants of theoretical department  seminar of PNPI
 for discussions of this work.

\end{document}